\documentstyle[prl,aps]{revtex}

\newcommand{\be}{\begin{equation}}
\newcommand{\ee}{\end{equation}}
\newcommand{\bea}{\begin{eqnarray}}
\newcommand{\eea}{\end{eqnarray}}
\renewcommand{\Im}{{\rm Im}\,}
\renewcommand{\Re}{{\rm Re}\,}

\begin{document}

\title{
Quantum and Thermal Depinning of a String from a Linear Defect
}
\author{Mikhail A.~Skvortsov}
\address{
L.~D.~Landau Institute for Theoretical Physics,
117940 Moscow, Russia\\
and Theoretische Physik, ETH-H\"onggerberg,
CH-8093, Z\"urich, Switzerland}

\maketitle

\begin{abstract}
The problem of a massive elastic string depinning from a linear defect
under the action of a small driving force is considered.
To exponential accuracy the decay rate is calculated with the help of
the instanton method; then, fluctuations of the quasiclassical solution
are taken into account to determine the preexponential factor.
The decay rate exhibits a kind of first order transition from quantum
tunneling to thermal activation with vanishing crossover region.
The model may be applied to describe nucleation in 2-dimensional first
order quantum phase transitions.
\end{abstract}



\section{ Introduction }

Macroscopic tunneling phenomena always attracted theorists' attention
since it is the field where classical and quantum physics meet each
other.  It governs the low temperature decay of metastable states of
physical systems with many degrees of freedom, separated from the
neighboring states by high potential barriers; at high temperatures
the metastable state decays via thermal activation.  Several such
systems were studied, including motion of dislocations across Peierls
barriers \cite{PP,IM}, tunneling with strong friction \cite{LO}, or
quantum breaking of an elastic string \cite{LSY}.

In the present paper the problem of quantum and thermal depinning of a
massive elastic string trapped in a linear defect and subject to a
small driving force is considered.  The interest in one-dimensional
manifold tunneling has revived last years as it describes the creep of
vortices trapped by columnar defects or plane twin boundaries
\cite{Review}.  Unlike thermal activation, which
for nondisspative Hamiltonian systems
is independent of the dynamic properties of the string \cite{SG},
quantum tunneling is strongly
dependent on the dynamics, which for the case of vortices in HTSC can
be either dissipative or governed by a Hall force \cite{Review}.  The
tunneling problem of a massive string therefore cannot be directly
applied to the vortex creep but its study is important since it allows
for an exact analytical solution.  Massive string depinning from a
linear defect describes the nucleation of a new phase in the vicinity
of a first order quantum phase transition in 2 dimensions.  The
boundary of the two phases may be considered as a string, the
difference in chemical potentials of the phases plays the role of a
driving force.

The problem is solved quasiclassically and exact solutions are
obtained for the whole temperature range in the limit of small driving
force.  Then, fluctuations of the quasiclassical solutions are taken
into account to determine the preexponentials.  The special feature of
the problem considered is the presence of a ``first order'' transition
between the quantum tunneling and thermal activation regimes that
should be contrasted with a smooth crossover between them in other
related problems \cite{PP,IM,Affleck}.

\section{ Model and Summary }

Consider a massive elastic string on a plane pinned by a linear
defect. The dynamics of the string is governed by the Lagrangian
\be
\label{Largangian}
{\cal L}[u(x,t)] =
\int dx\left[\frac{\rho}{2}\dot{u}^2
-\frac{\rho c^2}{2} (\partial_x u)^2 - V(u) \right].
\ee
Here $\rho$ is the mass density of the string, $c$ is the speed of
sound, and $u$ is the displacement perpendicular to the string.
The string is directed along the $x$-axis.
In the presence of a driving force, $V(u)$ can be represented as
\be
\label{V(u)}
V(u)=V_0(u)-Fu,
\ee
where $V_0(u)$ is a potential of a single pinning line of size $u_0$:
$V_0(0)=0$ and $V_0(|u|\gg u_0)=E_0$.
The driving force $F$ renders the states localized near $u=0$
metastable and the string depins and leaves the defect
in a finite time $\Gamma^{-1}$.
The lifetime $\Gamma^{-1}$ will be obtained analytically
for the whole temperature range in the limit of small $F$.

At zero temperature the string has no energy to overcome the potential
barrier that separates the states with $u=0$ and with $u\ge E_0/F$
where the barrier disappears. In this case the decay rate
is determined by macroscopic quantum tunneling.
To study the tunneling process one has to
rewrite the action in imaginary time (for temperature $T$):
\be
\label{action}
A[u(x,\tau)] = \int_{-\hbar/2T}^{\hbar/2T}d\tau
\int dx\left[\frac{\rho}{2}\dot{u}^2
+\frac{\rho c^2}{2}(\partial_x u)^2+V(u)\right],
\ee
and find the saddle point solution $\tilde u(x,\tau)$ for this action
that can be obtained from variation of (\ref{action}).
Such a solution of a Euclidean field theory is commonly referred to
as an instanton \cite{Raj}.
Then, to exponential accuracy
the lifetime $\Gamma^{-1}\propto \exp(-A[\tilde u]/\hbar)$.

The action can be easily estimated as follows \cite{Dyakonov,LSY}.
Let $\tau$ be the tunneling time.
Then the mass involved in the tunneling process can be estimated as
$M\sim \rho c\tau$. We can describe the collective tunneling process
as a tunneling of one particle
of mass $M$ through a potential barrier of height $E_0c\tau$.
The tunneling time can be obtained from the condition that
the kinetic energy $M(E_0/F)^2/\tau^2$ should be of the order
of the potential energy $E_0c\tau$.
This gives the estimate $\tau^2\sim \rho E_0/F^2$ and the action
\be
\label{estimQ}
A_Q \sim \frac{\rho cE_0^2}{F^2}.
\ee
Such a quasiclassical treatment is valid provided that
$A_Q/\hbar\gg1$.
This criterion can be expressed as
\be
\label{cond1}
\frac{\hbar}{\rho cu_0^2} \alpha^2 \ll 1,
\ee
where $\alpha=F/F_0$ and $F_0\sim E_0/u_0$ is the typical value
of the force that keeps the string in the well.
$F_0$ is also of the order of the critical
driving force $F_c$ at which the potential barrier disappears.
The following consideration will be constrained to the limit of small
$\alpha \ll 1$.

At high temperatures when the periodicity $\hbar/T$ along the
$\tau$-axis becomes smaller than the size
$\sim(\rho E_0)^{1/2}/F$ of the quantum
instanton, the saddle point solution $\tilde u(x)$ does not
depend on $\tau$ any more and the escape rate is determined
by pure thermal activation.
In this case $\Gamma^{-1}\propto \exp(-\Delta/T)$,
where $\Delta$ is the energy of the optimal configuration
$\tilde u(x)$ that extremizes (\ref{action}).
Note that it is independent of the dynamics since the thermal
saddle point solution does not depend on imaginary time.
The energy barrier $\Delta$ can be estimated
in a way similar to the one used for the quantum case.
Dimensional analysis yields
\be
\label{estimT}
\Delta \sim \frac{(\rho c^2 E_0^3)^{1/2}}{F}.
\ee

The activation energy diverges at zero driving force.
Such a behavior should be contrasted with the thermal kink--antikink
nucleation in the models with several degenerate minima (Sine Gordon,
$\phi^4$ theories) \cite{PP,IM}, where in the absence of a driving
force the activation barrier remains finite and is twice as
large as the kink rest energy $E_K$.
In the present case, the single minimum of $V_0(u)$ is nondegenerate,
resulting in large $F$-dependent size of the instanton and
divergent $\Delta$.

There exist a temperature $T_c$ when the probabilities of quantum and
thermal instanton nucleations become equal,
\be
\label{}
T_c \sim \hbar\frac\Delta{A_Q} \sim \frac{\hbar F}{(\rho E_0)^{1/2}}.
\ee
Usually, this temperature indicates a crossover
from quantum tunneling to thermal activation.
This is the case for single particle tunneling \cite{Affleck},
tunnelling of a string between two degenerate minima when a small
driving force lifting the degeneracy is applied \cite{PP,IM},
and quantum breaking of an elastic string \cite{LSY}.
In the present problem the situation is different.
It will be shown that for $T<T_c$ the string escapes
due to quantum tunneling, temperature corrections to the action
being exponentially small for $\alpha\ll1$.
For $T>T_c$ the lifetime is determined by pure thermal activation.
Hence, at $T=T_c$ the system exhibits a kind
of first-order phase transition from the quantum
to the thermal regime and no crossover region in the
conventional sense is obtained.
Such a behavior is a consequence of the coexistence of
{\it two qualitatively different}
quantum and thermal instantons in the vicinity of $T_c$.
The total $\Gamma^{-1} = (\Gamma_Q + \Gamma_T)^{-1}$
is equal to $\Gamma_Q^{-1}$ for $T<T_c$ and to
$\Gamma_T^{-1}$ for $T>T_c$.
On the other hand, in a usual situation there exist only {\it one}
saddle point solution of (\ref{action}) that describes quantum
tunneling at low temperatures, pure activation at high temperatures,
and crosses over between them at $T\sim T_c$.

\section{ Quasiclassical Analysis }

In this section the saddle point solutions for the action
(\ref{action}) are obtained in the limit of small driving force.
For further use I want to specify some quantities
defined above by order of magnitude estimates.
We will see below that in the limit of small $F$ the quasiclassical
results for the nucleation rate are independent of the certain
features of $V_0(u)$ at the scale $u_0$.
The only relevant parameter is the depth $E_0$ of the potential.
The other characteristic of $V_0(u)$ that determines
instantons interaction and the strength of quantum fluctuations
is the curvature at the origin $V_0''(0)$.
Instead of it I define the width of the pinning potential
according to
\be
u_0^2 = \frac{2E_0}{V_0''(0)}.
\ee
The ratio of the driving force to the pinning force is the small
parameter in the problem:
\be
\alpha = \frac{Fu_0}{E_0} =
\left[ \frac{2F^2}{E_0 V_0''(0)} \right]^{1/2}.
\ee

In the following I will work in the units $\rho=c=1$.
One has to rescale $x\to(\rho c^2)^{-1/2}x, \tau\to\rho^{-1/2}\tau$
to return to dimensional units.
The classical equation of motion obtained from
the action (\ref{action}) is
\be
(\partial_x^2 + \partial_\tau^2) u=V'(u).
\label{sp}
\ee
The function $u(x,\tau)$ is periodic in $\tau$ with
the period $\hbar/T$ and vanishes at $|x|\to\infty$.
Eq.~(\ref{sp}) defines a 2D nonlinear electrostatics problem,
the charge density being dependent on the electrical potential $u$.

The solution of Eq.~(\ref{sp}) behaves essentially different
for $u\ll u_0$ and for $u\gg u_0$ (see Eq.~(\ref{V(u)})
for the definition of $V(u)$).
Let us define a curve $\Gamma$ on the $(x,\tau)$ plane where $u=u_0$.
It separates the plane into inner and outer domains
$\cal B$ and $\bar{\cal B}$.
In the outer domain $\bar{\cal B}$, $u<u_0$
that corresponds to the string lying in the pinning well $V_0(u)$.
Inside the domain $\cal B$, $u>u_0$ and the string
is under the barrier.

The key point in the determination of $u(x,\tau)$ from the nonlinear
Eq.~(\ref{sp}) is to obtain the solution in the outer
domain $\bar{\cal B}$ and to extract the boundary condition
for the internal region of the instanton.
The size of the instanton ($R$ for the quantum instanton
(\ref{qu}), or $L$ for the thermal instanton (\ref{th});
see Fig.~1 for geometry clarifying)
is of the order of $E_0^{1/2}/F$ as estimated above.
In the external region $\bar{\cal B}$ of the instanton,
$u$ rapidly falls off to zero at a distance $a$ that can be estimated
from Eq.~(\ref{sp}) as $a\simeq [V''(0)]^{-1/2} \sim \alpha R$.

Since $a$ is much smaller than the boundary curvature radius
($R$ for the quantum instanton,
or $\infty$ for the thermal instanton),
one can neglect the curvature in the Laplace operator
in a layer of width $a$ along the boundary $\Gamma$ of the instanton
and write simply
$\nabla^2 \simeq \frac{\partial^2}{\partial n^2}$,
where ${\bf n}$ is the outward
normal to the boundary of the instanton.
Then Eq.~(\ref{sp}) becomes the equation of motion of a classical
particle in an inverted potential $-V(u)$:
\be
\frac{d^2u}{dn^2}=V'(u).
\ee
The energy ${\cal E}=\frac12\left(\frac{du}{dn}\right)^2-V(u)$
is an integral of the motion.
Its value should be obtained from the condition at infinity,
that gives ${\cal E}=0$.
Applied to $u=u_0$, the equation ${\cal E}=0$ gives the boundary
condition for the interior problem:
\be
\label{bc}
\left.\frac{du}{dn}\right|_\Gamma=-\sqrt{2E_0}.
\ee

The other condition is $u|_\Gamma=u_0$.
Thus, we arrive at the following problem:
\bea
\nabla^2 u &=& -F, \label{pe}\\
u|_\Gamma &=& u_0, \label{ug}\\
\left.\frac{du}{dn}\right|_\Gamma &=& -\sqrt{2E_0}. \label{dug}
\label{problem}
\eea

Eqs.~(\ref{pe}) and (\ref{ug}) define a Dirichlet problem with
a unique solution for a given contour $\Gamma$.
The solution of Eq.~(\ref{pe}) with the boundary
condition (\ref{dug}) (von Neumann problem) is unique as well.
The problem (\ref{pe})--(\ref{dug}) is therefore strongly
overdetermined and cannot be solved for a general contour $\Gamma$.

There are only two regions for which the solution of Eq.~(\ref{pe})
satisfies both boundary conditions simultaneously.
They are (i) the circle and (ii) the strip (see Fig.~1).

(i) Circle of radius $R$. The solution is
\be
\label{qu}
u=u_0 + \frac F4 (R^2-r^2),
\ee
where $r^2=x^2+\tau^2$. The radius can be obtained from (\ref{bc}):
\be
\label{R}
R^2=\frac{8E_0}{F^2}.
\ee
This is the quantum instanton.

(ii) Strip of width $2L$ parallel to the $\tau$-axis. Then
\be
\label{th}
u=u_0 + \frac F2 (L^2-x^2)
\ee
with
\be
\label{L}
L^2=\frac{2E_0}{F^2}.
\ee
This solution is time-independent and holds at large temperatures.

The solutions are shown schematically in Fig.~2.

To compute the action one has to know the solution of Eq.~(\ref{sp})
everywhere. The contribution of the boundary layer to the action
is not universal and depends on the features of the pinning
potential $V_0(u)$ but is of order $\alpha$ when compared
to the contribution from the interior region.
Thus, to leading order in $\alpha$,
the action for the circle is (here and below dimensional units are restored)
\be
\label{aqu}
A_{Q}=4\pi\frac{\rho cE_0^2}{F^2},
\ee
whereas for the strip (temperature $T$) the result reads
\be
\label{ath}
A_{T}(T)=\hbar\frac{\Delta_0}T,
\ee
with $\Delta_0$ the energy cost of the optimal tongue (\ref{th}),
\be
\Delta_0 = \frac{2^{5/2}\rho^{1/2}c E_0^{3/2}}{3F}.
\ee

Quantum and thermal actions become equal at
\be
\label{tc}
T_c = \frac{2^{1/2}}{3\pi} \frac{\hbar F}{(\rho E_0)^{1/2}}.
\ee
The strip (\ref{th}) is a solution for all $T$.
It will be shown in the next section that this solution becomes
unstable below some temperature $T_0 = 0.9 T_c$.
The circle solution exists as long as its diameter $2R$ is less than
the periodicity $\hbar/T$ in $\tau$. The quantum solution then
ceases to exist at
\be
\label{t1}
T_1 = \frac{3\pi}8 T_c.
\ee
The dependence of the actions on temperature for the two solutions is
illustrated in Fig.~3(a).

The lifetime $\Gamma^{-1}$ is determined by the solution with the
minimal action. To exponential accuracy,
\be
\label{lnG}
\ln\Gamma^{-1}\propto
\cases{
A_Q/\hbar  &for $T<T_c$,\cr
\Delta_0/T &for $T>T_c$.\cr}
\ee
This answer holds for all temperatures but not too close to $T_c$.
In a very narrow region around $T_c$, where
$|1-T/T_c| < T_c/\Delta \sim \alpha^2 \frac{\hbar}{\rho c u_0^2} \ll1$
(cf. (\ref{cond1})),
both solutions contribute to $\Gamma^{-1}$
according to the formula $\Gamma^{-1} = (\Gamma_Q + \Gamma_T)^{-1}$.
Note that the last equation is not a bridging formula
but an exact expression up to exponentially small corrections.
This is a consequence of the coexistence of two types of instantons
close to $T_c$.

It is also instructive to study the temperature dependence
of the quantum tunneling probability.
It appears to be very small due to the
exponential decay of $u$ at large distances.
Actually, the distance between the boundary of the
instanton (\ref{qu}) and its image is
$\delta\tau = \hbar/T - \hbar/T_1$.
Then according to Eq.~(\ref{sp}) the overlap of the
exponentially decaying tails
of the instanton and its image is as small as
$u_0\exp[-(V''(0)/\rho)^{1/2} \delta\tau]$, resulting in
\be
\frac{\delta A_Q(T)}{A_Q} \sim
\exp \left[
        -\left(
            \frac{2E_0}{\rho u_0 ^2}
        \right)^{1/2}
    \left(
        \frac\hbar{T} - \frac\hbar{T_1}
    \right)
\right],
\ee
In the vicinity of $T_1$,
\be
\label{temp_corr}
\frac{\delta A_Q(T)}{A_Q} \sim
\exp \left[
    -\frac8{\alpha}
    \left(
        1 - \frac{T}{T_1}
    \right)
\right].
\ee

\section{Fluctuations}

Let us now take fluctuations near the classical instantons
into account and determine the preexponential factors in (\ref{lnG}).
The starting point is the expression for the lifetime $\Gamma^{-1}$
\cite{Langer,LO}
\be
\hbar\Gamma=2T\frac{\Im Z}{\Re Z},
\ee
where $Z$ is the partition function
\be
Z=\int Du(x,\tau) \exp\left(-\frac{A[u]}{\hbar}\right).
\ee
The neighborhood of the saddle point solution $\tilde u(x,\tau)$
of the equation
$\delta A[u]/\delta u = 0$ gives the main contribution to the
imaginary part of $Z$. I expand $u$ near $\tilde u$ according to
\be
u(x,\tau)=\tilde u(x,\tau) + \sum_\alpha C_\alpha u_\alpha(x,\tau),
\ee
where $u_\alpha(x,\tau)$ are the eigenfunctions of the operator
$\delta^2 A[u]/\delta u^2$ :
\be
\label{eigen}
\left( \delta^2 A[u]/\delta u^2 \right)_{\tilde u} u_\alpha(x,\tau) =
\lambda_\alpha u_\alpha(x,\tau).
\ee

The lowest eigenvalue $\lambda_0$ is negative, resulting
in the imaginary contribution to the partition function.
$\lambda_1=0$ is also an eigenvalue
of (\ref{eigen}) related to the zero mode with respect to translation.
For the thermal instanton (\ref{th}), $\lambda_1=0$ is nondegenerate
and $u_1(x)\propto\partial_x\tilde u$ corresponds to the zero mode
along $x$-axis,
while for the quantum instanton (\ref{qu}),
the zero level is twofold degenerate
since the location of the instanton
on the $x,\tau$ plane depends on both coordinates.

Applying standard methods \cite{ZL,CC,LO} for zero modes and
the negative eigenvalues we arrive at the following expressions:
The lifetime per unit length for thermal activation is
\be
\label{Gth}
\frac\Gamma L =
\left[ \frac{T}{2\pi\hbar^2}
       \int dx \left( \frac{\partial\tilde u}{\partial x} \right)^2
\right]^{1/2}
\left|
    \frac{\det'
        \left(
            \delta^2 A[u]/\delta u^2
        \right)_{u=\tilde u}
    }{\det
        \left(
            \delta^2 A[u]/\delta u^2
        \right)_{u=0}
    }
\right|^{-1/2}
\exp \left( -\frac{A[\tilde u]}{\hbar} \right).
\ee
The decay rate for quantum tunneling reads
\be
\label{Gqu}
\frac\Gamma L =
\frac1{2\pi\hbar}
\left[
    \int\!\!\int dx d\tau
        \left( \frac{\partial\tilde u}{\partial x} \right)^2
\right]^{1/2}
\left[
    \int\!\!\int dx d\tau
        \left( \frac{\partial\tilde u}{\partial\tau} \right)^2
\right]^{1/2}
\left|
    \frac{\det''
        \left(
            \delta^2 A[u]/\delta u^2
        \right)_{u=\tilde u}
    }{\det
        \left(
            \delta^2 A[u]/\delta u^2
        \right)_{u=0}
    }
\right|^{-1/2}
\exp \left( -\frac{A[\tilde u]}{\hbar} \right).
\ee
Here $\det'$ and $\det''$ denote the determinants
with the zero eigenvalues omitted.

\subsection{ High-temperature instanton}

For the high temperature $\tau$-independent solution (\ref{th})
(I use the units $\hbar=\rho=c=1$),
\be
\label{d2Adu2}
\frac{\delta^2A}{\delta u^2} =
-\partial_\tau^2 - \partial_x^2 + U(x);
\ee
\be
U(x) = \frac{\partial^2 V(\tilde u(x))}{\partial u^2}.
\ee

The eigenvalues of the operator (\ref{d2Adu2}) are given by
\be
\label{lambda_mn}
\lambda_{mn}=\epsilon_{m} + (2\pi T n)^2,
\ee
where $n$ is an integer and $\lbrace \epsilon_m \rbrace$ is
the spectrum of the one dimensional Schr\"odinger equation
\be
\label{SE}
[-\partial_x^2 + U(x)]\psi_m(x) = \epsilon_m \psi_m(x).
\ee

The potential $U(x)$ is illustrated schematically in Fig.~4.
$L\propto \alpha^{-1}$ is the semiwidth of the instanton
strip (\ref{th}), the width of the minimum of $U(x)$
near $|x|=L$, $a\sim L\alpha$,
is independent of $\alpha$;  $U(|x|\gg L)=\omega^2\equiv V_0''(0)$
is the oscillator frequency for the pinning potential $V_0(u)$.
The form of the potential near $|x|=L$ depends on the detailed
features of $V_0(u)$.
For $\alpha\to 0$, $L\to\infty$ and the potential well $U(x)$
becomes deep in the sense that as many as $\frac4\pi \alpha^{-1}$
discrete levels exist in this potential.
Excited states can be obtained within
the quasiclassical approximation.
It is possible also to
obtain the low lying levels in the limit of small $\alpha$.

Let us define $\gamma$ as a function of the continuous
parameter $\epsilon$ as
\be
\gamma(\epsilon)=\partial_x\ln\psi(L),
\ee
where $\psi(x)$ is the solution of Eq.~(\ref{SE}) with
energy $\epsilon$ that vanishes at $x\to\infty$.
Once the function $\gamma(\epsilon)$ is known,
one can obtain the discrete spectrum in the following way.
For $|x|<L$, the solutions of Eq.~(\ref{SE})
are given by $\psi_\epsilon(x) = \cos(\sqrt{\epsilon}x)$ and
$\psi_\epsilon(x) = \sin(\sqrt{\epsilon}x)$
for even and odd states respectively. The spectrum then can
be obtained from the continuity of the
logarithmic derivative of $\psi_\epsilon(x)$ at $x=L$:
\be
\label{spectrum}
\partial_x\ln\psi_\epsilon(L) = \gamma(\epsilon).
\ee

The function $\gamma(\epsilon)$ is not universal and depends
on the shape of the minimum of $U(x)$,
that is on the details of $V_0(u)$.
It appears however that $\gamma(0)$
is independent of the pinning potential.
We may find $\gamma(0)$ by using our knowledge that
the first excited level has zero energy.
For $|x|<L$ its wave function linearly depends on $x$ yielding
\be
\gamma(0)=1/L.
\ee

The low lying states can be obtained from Eq.~(\ref{spectrum})
with the right hand side set equal to $\gamma(0)$ thus neglecting
the difference between $\gamma(\epsilon)$ and $\gamma(0)$
for small $\epsilon$.
Then one immediately obtains the implicit equations for the spectrum:
\be
\label{even}
\sqrt\epsilon L \tan(\sqrt\epsilon L) = -1
\ee
for even states and
\be
\label{odd}
\sqrt\epsilon L \cot(\sqrt\epsilon L) = 1
\ee
for odd states.

The approximation $\gamma(\epsilon) \simeq \gamma(0)$ is valid for
small enough energies
$\epsilon \ll \epsilon_* \sim \gamma(0)/\gamma'(0)$.
One can estimate $\gamma'(0)$
as $\gamma(\omega^2)/\omega^2 \sim 1/\omega$,
yielding $\epsilon_* \sim \alpha^{-1}/L^2$.
On the other hand, the energy of the $n$-th excited level can
be estimated as $\epsilon_n \simeq (\pi n/(2L))^2$ for large $n$.
Therefore, Eqs.~(\ref{even}) and (\ref{odd}) correctly describe
as many as $\alpha^{-1/2}$ low lying levels.

The negative eigenvalue $\epsilon_0$ can be obtained from
Eq.~(\ref{even}) and is equal to
\be
\label{epsilon_0}
\epsilon_0=-\frac{\mu^2}{L^2},
\ee
where $\mu=1.19968$ is the solution of the equation $\mu\tanh\mu=1$.

The negative eigenvalue determines the temperature $T_0$
where $\lambda_{01}$ changes the sign leading to the instability of
the thermal instanton (\ref{th})
with respect to a small perturbation of its boundary.
Inserting (\ref{epsilon_0}) into Eq.~(\ref{lambda_mn}), I obtain
\be
\label{T0}
T_0 = \frac{\mu}{2\pi} \frac1L,
\ee
in dimensional units,
\be
\label{T_0}
T_0 = \frac{\mu}{2^{3/2}\pi} \frac{\hbar F}{(\rho E_0)^{1/2}}
    = 0.900 T_c.
\ee

Now we are in a position to calculate
the ratio of determinants in (\ref{Gth}).
To do it, one has to compute the products
over $n$ \cite{IM}, resulting in
\be
\label{det_ratio}
\left|
    \frac{\det'
        \left(
            \delta^2 A[u]/\delta u^2
        \right)_{u=\tilde u}
    }{\det
        \left(
            \delta^2 A[u]/\delta u^2
        \right)_{u=0}
    }
\right|^{-1/2}
=
\frac{T}{2\sin\frac{\pi T_0}{T}}
    e^{-\Delta_1/T}
    \chi \left( \frac{T}{T_0} \right),
\ee
with the quantum correction to the barrier height,
\be
\label{Delta1}
\Delta_1 =
\frac12\sum_{m=2}^M\epsilon_m^{1/2} +
\int_{-\infty}^\infty
     \frac{dq}{4\pi}\frac{\partial\delta(q)}{\partial q}
     \epsilon(q)^{1/2}.
\ee
Here $\delta(q)$ is the phase shift of the states
in the continuous spectrum, $\epsilon(q)=\omega^2+q^2$,
and $M$ labels the last discrete state.
The function $\chi$ is defined according to
\be
\label{chi}
\ln\chi \left( \frac{T}{T_0} \right) =
        - \sum_{m=2}^M \ln(1-e^{-\epsilon_m^{1/2}/T}) -
        \int_{-\infty}^\infty
            \frac{dq}{2\pi}\frac{\partial\delta(q)}{\partial q}
            \ln(1-e^{-\epsilon(q)^{1/2}/T}).
\ee

At large $q$, $\delta(q)$ can be calculated perturbatively \cite{LL3},
\be
\delta(q) = -\frac1{2q} \int [U(x)-\omega^2] dx = \frac{\omega^2L}{q},
\ee
leading to a logarithmic divergence of the integral
over continuous spectrum in (\ref{Delta1})
that should be cut off at some wave vector $\Lambda$:
\be
\Delta_1 = -\frac{\omega^2L}{4\pi}\ln\frac\Lambda\omega.
\ee
The ultraviolet cutoff $\Lambda$ is a new parameter that should
be introduced for the problem described by the Lagrangian
(\ref{Largangian}) to be well defined.
The value of $\Lambda$ remains undefined in the model
in question. Physically, it is determined by the length scale at which
the elastic approximation of the bending energy in (\ref{Largangian})
breaks down. Then $\Lambda$ may be estimated
as the inverse core radius of the linear manifold considered.

$\chi(T/T_c)$ is a complicated function of temperature.  For $T$
close to $T_0$, only the lowest level with positive energy
$\epsilon_2$ is excited.  Its energy can be obtained from
Eq.~(\ref{even}): $\epsilon_2=(2.79839/L)^2$, resulting in
\be
\chi(x\sim1) \simeq 1 + \exp\left(-\frac{14.7}x \right).
\ee

For intermediate temperatures, $T_0\ll T\ll T_0/\alpha \sim\omega$,
many levels contribute to the sum in (\ref{chi}). In this limit,
$\epsilon_m \sim (\pi m/(2L))^2$ and
\be
\label{chi2}
\chi(x)= \exp \left( \frac\mu6x \right).
\ee

For $T\gg T_0$, many levels are excited and the quasiclassical
approach can be applied. The sum over the states
in Eq.~(\ref{chi}) can be expressed via an integral over phase space:
\bea
\ln\chi \left( \frac{T}{T_0} \right) &\simeq&
    - 2 \ln ( 1-e^{-T_0/T} )
    + \frac1{2\pi} \int d x \, d p
        \left[
            \ln ( 1-e^{-\sqrt{\omega^2+p^2}/T} )
            - \ln ( 1-e^{-\sqrt{U(x)+p^2}/T} )
        \right] \nonumber          \\
&=&
    - 2 \ln ( 1-e^{-T_0/T} )
    + \frac{2L}\pi \int_0^\infty d p
        \left[
            \ln ( 1-e^{-\sqrt{\omega^2+p^2}/T} )
            - \ln ( 1-e^{-p/T} )
        \right].
\label{qc1}
\eea
The first term in (\ref{qc1}) accounts for the two omitted states
in Eq.~(\ref{chi}). For $T\ll T_0/\alpha$, Eq.~(\ref{qc1}) reduces
to Eq.~(\ref{chi2}). For $T\gg T_0/\alpha$, one finds
\be
\label{chi3}
\chi(x) = \frac1{x^2} \exp\left( \frac2\alpha \right).
\ee

Substituting (\ref{th}) and (\ref{det_ratio}) into (\ref{Gth}),
we obtain
\be
\label{Gth1}
\frac{\Gamma}L = B(T) \exp \left( -\frac{\Delta}{T} \right).
\ee
$\Delta=\Delta_0 + \Delta_1$ and $B(T)$ in conventional units are
given by the following expressions:
\be
\label{res1}
\Delta =
\frac{2^{5/2}\rho^{1/2}c E_0^{3/2}}{3F}
\left(
    1 - \frac{3\hbar}{8\pi\rho c u_0^2}
    \ln\left[
        \Lambda u_0
        \left(
            \frac{\rho c^2}{E_0}
        \right)^{1/2}
    \right]
\right),
\ee
\be
B(T) = \frac1{2^{1/4}3^{1/2}\pi^{1/2}}
       \frac{\rho^{1/4}E_0^{3/4}}
           {\hbar^2c^{1/2}F^{1/2}}
       \frac{T^{3/2}} {\sin\frac{\pi T_0}T}
       \chi \left( \frac{T}{T_0} \right).
\ee

\subsection{ Quantum instanton }

Fluctuations of the quantum instanton (\ref{qu}) are governed
by the following Hamiltonian
\be
\label{d2Adu2-qu}
\frac{\delta^2A}{\delta u^2} =
-\partial_\tau^2 - \partial_x^2 + U(r);
\ee
where $U(r)$ is cylindrically symmetrical,
\be
\label{U-qu}
U(r) = \frac{\partial^2 V(\tilde u(r))}{\partial u^2}.
\ee
The potential $U(r)$ looks very similar to the one-dimensional
potential $U(x\geq0)$ illustrated in Fig.~4.

The ratio of determinants in (\ref{Gqu}) can be expressed in terms
of the spectrum as
\be
\label{det-ratio}
\left|
    \frac{\det''
        \left(
            \delta^2 A[u]/\delta u^2
        \right)_{u=\tilde u}
    }{\det
        \left(
            \delta^2 A[u]/\delta u^2
        \right)_{u=0}
    }
\right|^{-1/2}
=
\exp\left\{
    -\frac12
    \mathop{{\sum}''}_m \ln|\epsilon_m|
    +\frac12 \sum_n \ln|\epsilon_n|
\right\},
\ee
where the first sum is taken over the states in the potential
(\ref{U-qu}) and the second in the potential $U(r)=\omega^2$.

We can estimate (\ref{det-ratio}) within the quasiclassical approach
by taking advantage of the large number
of bound states in the potential (\ref{U-qu}) for small $\alpha$.
According to the quasiclassical quantization rule,
each state occupies the volume $(2\pi)^2$ in phase space and
(\ref{det-ratio}) reduces to
\be
\epsilon
\exp\left\{
    -\frac12 \frac1{(2\pi)^2}
    \int d^2r \, d^2p
    \left[
        \ln\sqrt{U(r)+p^2} - \ln\sqrt{\omega^2+p^2}
    \right]
\right\}
=
\epsilon
\exp\left\{
    \frac{R^2}8 \int_0^\infty p dp
        \ln\frac{\omega^2+p^2}{p^2}
\right\}.
\ee
The integral over the momentum is logarithmically divergent,
leading to
\be
\label{det-r}
\epsilon
\exp\left(
    \frac{R^2\omega^2}8 \ln \frac\Lambda\omega
\right).
\ee

Inserting (\ref{det-r}) in (\ref{Gqu}), and using (\ref{qu}) rewritten
in conventional units, I obtain the final result
\be
\label{Gqu1}
\frac{\Gamma}L = \frac{E_0}\hbar
\exp \left( -\frac{A}{\hbar} \right),
\ee
\be
\label{res2}
A = 4\pi\frac{\rho cE_0^2}{F^2}
\left(
    1 - \frac{\hbar}{2\pi\rho c u_0^2}
    \ln\left[
        \Lambda u_0
        \left(
            \frac{\rho c^2}{E_0}
        \right)^{1/2}
    \right]
\right).
\ee

The results (\ref{res1}) and (\ref{res2}) are valid provided that the
fluctuation contribution is less than the classical one.
This provides the condition
\be
\label{cond2}
\frac{\hbar}{2\pi\rho c u_0^2}
\ln\left[
    \Lambda u_0
    \left(
        \frac{\rho c^2}{E_0}
    \right)^{1/2}
\right] \ll 1.
\ee
This is a much stronger condition than that of (\ref{cond1}).
When (\ref{cond2}) fails then one has to take unharmonic fluctuations
into account that will renormalize the numerical coefficient in front
of (\ref{res2}), whereas the estimate (\ref{estimQ}) holds as long
as the condition (\ref{cond1}) is fulfilled.

\section{ Conclusion }

The problem of a massive string depinning from a linear defect has
been solved in the limit of small driving force ($\alpha\ll1$)
allowing for an exact analytical solution
in the whole temperature range.
It has been shown that there exists a temperature region $T_0<T<T_1$
where the action (\ref{action}) has {\em two} different saddle point
solutions: (\ref{qu}) and (\ref{th}).
The actions for these solutions become equal at $T=T_c$.
For $T<T_c$, the lifetime is determined by quantum tunneling
(Eq.~(\ref{Gqu1})) with exponentially small temperature corrections
(\ref{temp_corr}).
For $T>T_c$, the lifetime is given by a pure activation
expression (\ref{Gth1}). At $T=T_c$ the system jumps from quantum to
classical behavior.
The dependence of $\ln \Gamma^{-1}$ vs.\ $T$ at constant
$F$ is shown in Fig.~3(a).
The possibility of such a first order transition between quantum
and thermal behavior was first discovered in Ref.~\cite{LK}.

It is worth emphasizing that the first order transition is obtained
for small $\alpha\ll1$.
As $\alpha$ increases, $T_0(\alpha)$ and $T_1(\alpha)$
come closer to each other and become equal at some $\alpha_c\sim1$,
that is, there exists a tricritical point where the first order
transition disappears. For $\alpha$ close but below $\alpha_c$,
the temperature corrections
to the quantum tunneling rate become large and the transition
at $T_c(\alpha)$ should be considered as a transition
from thermally assisted quantum tunneling
to pure activation.

Besides the transition at constant $F$, one can consider the
transition at constant $T$ by changing the applied force which
may be easier to achive experimentally.
$\ln \Gamma^{-1}$ as a function of $F$ is shown in Fig.~3(b).
For $F<F_c = \frac{3\pi}{2^{1/2}} \frac{T}\hbar (\rho E_0)^{1/2}$,
the metastable state decays via thermal activation with
$\ln\Gamma^{-1} \propto F^{-1}$, whereas for $F>F_c$,
the string escapes via quantum tunneling with
$\ln\Gamma^{-1} \propto F^{-2}$.

The problem in question is connected with the nucleation phenomenon
in 2-dimensional first order quantum phase transitions.
Consider a first order phase transition controlled
by some parameter $x$. The quantum nature of the transition
implies that it exists even at $T=0$.
One possible example is the liquid--solid transition
in 2D helium films.
Let $x>0$ correspond to phase I, and $x<0$ to phase II.
The phase boundary is a string that has mass density $\rho$ and
elasticity $\rho c^2 = \sigma$, where $\sigma$ is the boundary
line tension.
The difference in chemical potentials of the phases is equivalent to
the driving force $F=\mu_I-\mu_{II} \propto x$.
Suppose that phase II is prepared at $x>0$. Such a state is metastable
and is destroyed by the nucleation of phase I that can take place near
the edges of the 2D system or in the bulk.
Nucleation near the edges is exactly the problem of a massive string
depinning from a linear defect. In the present case, the edge of the
system plays the role of the pinning well with the depth $E_0=\sigma$.
It can be shown that nucleation always occurs near the edges since
the bulk tunneling action and activation barrier are larger than the
corresponding values for the edge: $A_{bulk}/A_{edge} = 3\pi/2^{5/2}$
and $\Delta_{bulk}/\Delta_{edge} = 16\sqrt2/15$.

Finally, the results (\ref{res1}) and (\ref{res2}) for the string on a
plane can be easily generalized to the string in $(1+d)$ dimensions.
The instantons (\ref{qu}) and (\ref{th}) remain the same since any
transverse displacement will cost extra bending energy.
The preexponentials will be different;
they will be multiplied by an additional factor
from transverse fluctuations. The main effect is the appearance
of a coefficient $d$ in front of the logarithm in formulae
(\ref{res1}) and (\ref{res2}).

\acknowledgments

I am grateful to M.V.Feigel'man for his support and fruitful
discussions, to G.Blatter, V.B.Geshkenbein, L.S.Levitov and
D.M.Kagan for helpful comments.
I express my gratitude to the Institute for Theoretical Physics,
ETH--Z\"urich, Switzerland, for hospitality.
This work was supported by the joint grant M6M300 from the ISF
and the Russian Government, by the RFBR grant 95-02-05720 and
by the Swiss National Foundation.

\newpage

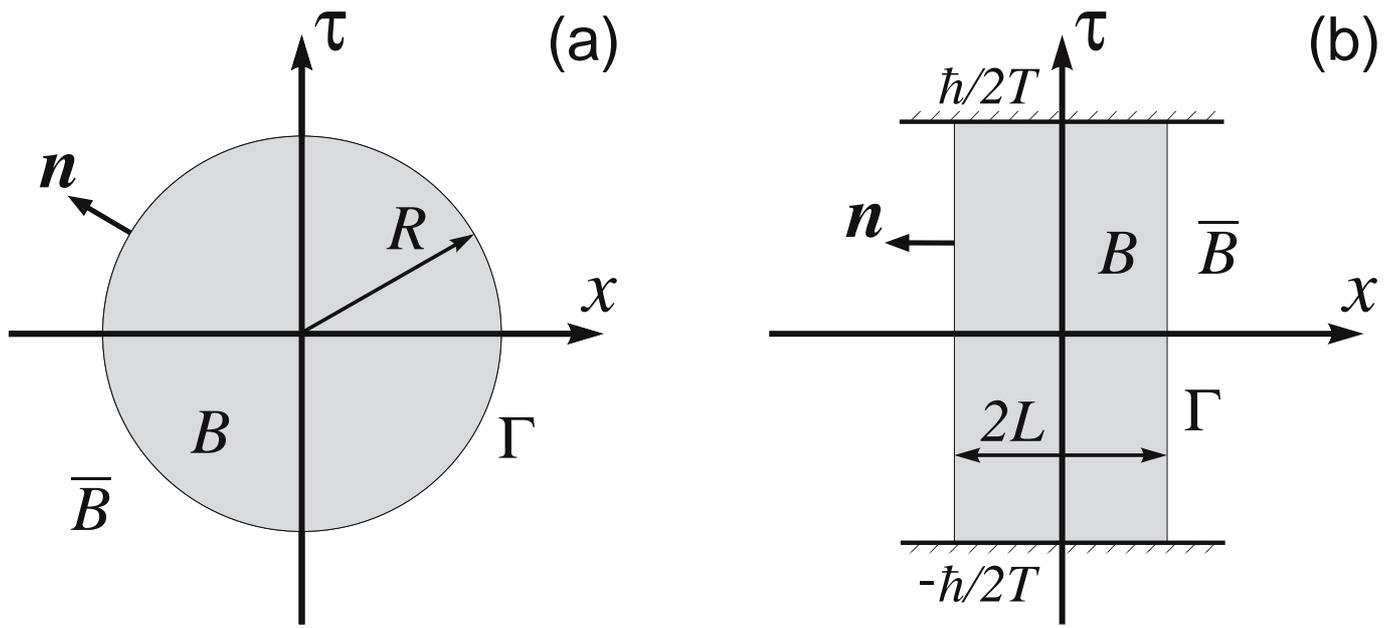
\begin{figure}
\caption{
The geometry of the quantum (a) and thermal (b) instantons.
}
\end{figure}

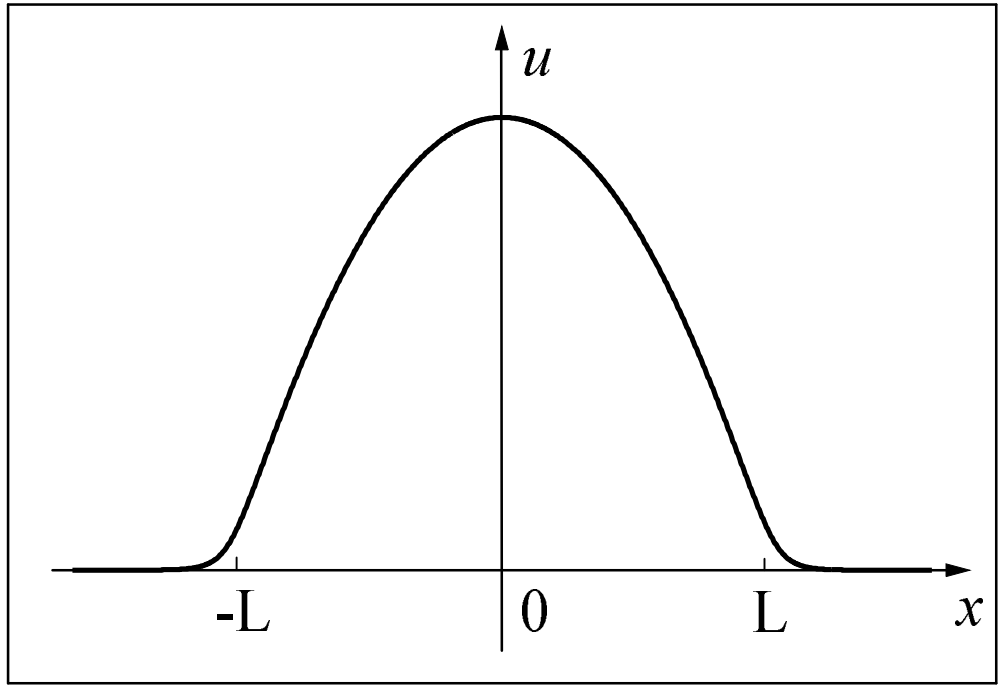
\begin{figure}
\caption{
The form of the thermal instanton. $u(x)$ is given by (18) for $|x|<L$
and exponentially decays for $|x|>L$.
The figure can also be considered as the radial dependence $u(r)$ of
the quantum instanton (16).
}
\end{figure}

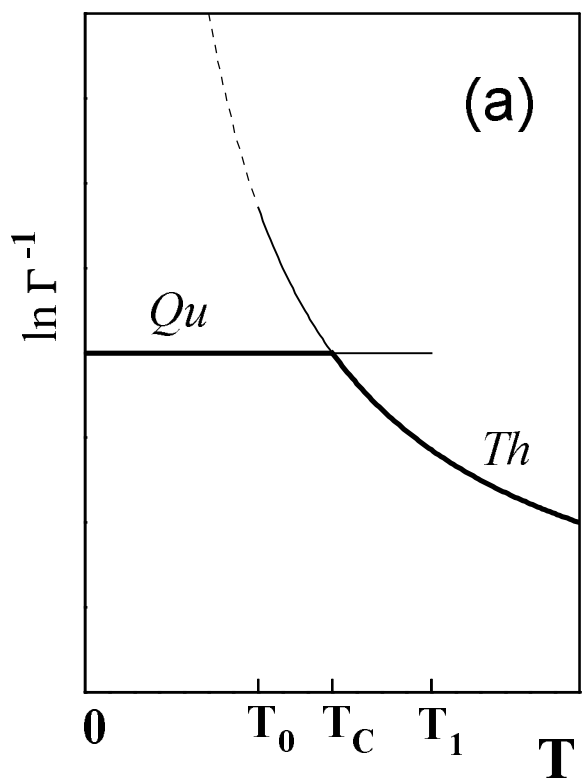
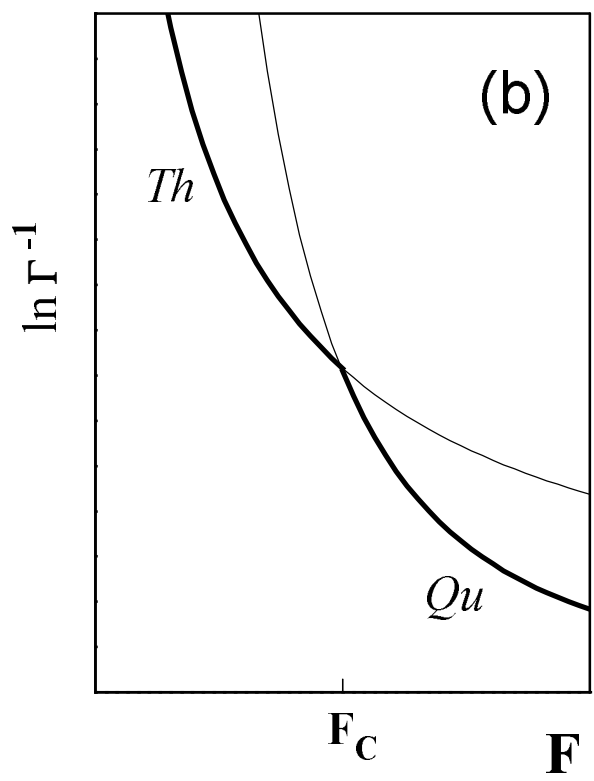
\begin{figure}
\caption{
(a) The dependence of the actions of the quantum (Qu) and thermal
(Th) instantons on temperature at constant $F$.
The decay rate exhibits a transition from
quantum tunneling to thermal activation at $T=T_c$.
(b) Decay rate vs.\ $F$ at constant $T$.
$F=F_c$ separates the quantum and thermal regimes.
}
\end{figure}

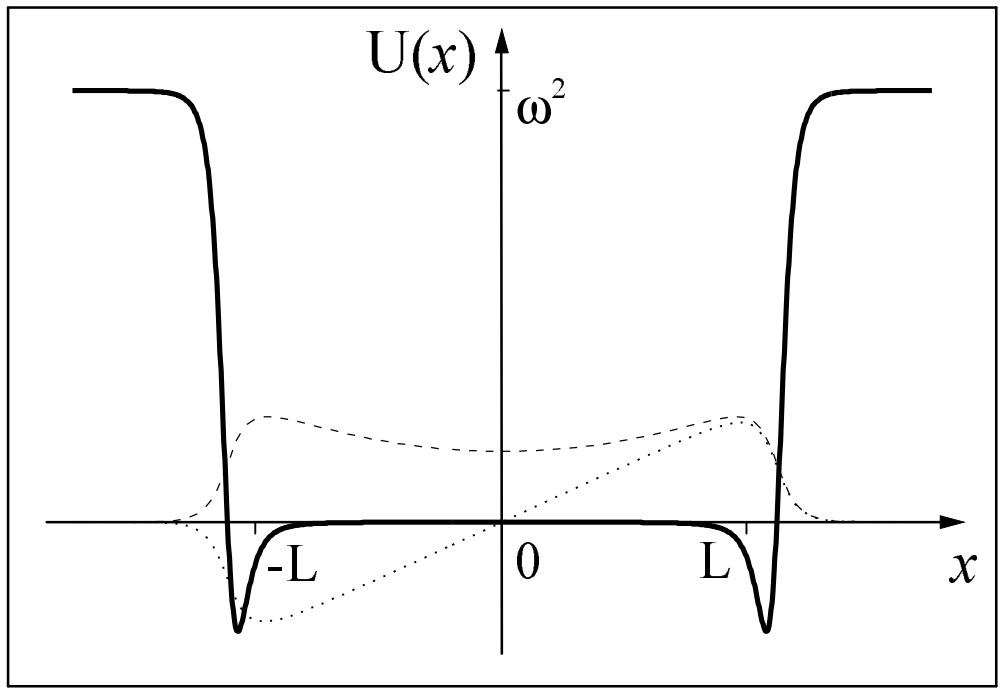
\begin{figure}
\caption{
Potential energy for the problem (37) linearized
in the vicinity of the thermal instanton.
Also shown are the wave functions of the ground (dashed line)
and the first excited (dotted line) states.
}
\end{figure}


\begin{references}

\bibitem{PP} B. V. Petukhov and V. L. Pokrovsky,
    Zh. Eksp. Teor. Fiz. {\bf 63}, 634 (1972)
    [Sov. Phys. -- JETP {\bf 36}, 336 (1973)].
\bibitem{IM} B. I. Ivlev and V. I. Mel'nikov, Phys. Rev. B {\bf 36},
    6889 (1987).
\bibitem{LO} A. I. Larkin and Yu. N. Ovchinnikov, Zh. Eksp. Teor. Fiz.
    {\bf 86}, 719 (1984) [Sov. Phys. -- JETP {\bf 59}, 420 (1984)].
\bibitem{LSY} L. S. Levitov, A. V. Shytov, and A. Yu. Yakovets,
    Phys. Rev. Lett. {\bf 75}, 370 (1995).
\bibitem{Review} For a review, see G. Blatter {\em et al},
    Rev. Mod. Phys. {\bf 66}, 1125 (1994).
\bibitem{SG} As an example of classical nucleation with dissipation,
    see F. Marchesoni, Phys. Rev. Lett. {\bf 73}, 2394 (1994)
    and references therein.
\bibitem{Affleck} I. Affleck, Phys. Rev. Lett. {\bf 46}, 338 (1981);
    U.Weiss, {\em Quantum Dissipative Systems}, World Scientific (1993).
\bibitem{Raj} R. Rajaraman, {\em Solitons and Instantons},
   North-Holland Publishing Company, Amsterdam/New York/Oxford (1982);
   S. Coleman, {\em Aspects of Symmetry}, Cambridge University Press (1985).
\bibitem{Dyakonov} M. I. Dyakonov,
    Fiz. Tver. Tela {\bf 29}, 2587 (1987).
\bibitem{Langer} J. S. Langer, Ann. Phys. {\bf 41}, 108 (1967).
\bibitem{ZL} I. Zittartz and J. S. Langer,
    Phys. Rev. {\bf 148}, 741 (1966).
\bibitem{CC} C. Callan and S. Coleman,
    Phys. Rev. D {\bf 16}, 1762 (1977).
\bibitem{LL3} L. D. Landau and E. M. Lifshitz,
    {\em Quantum Mechanics}, Course in Theoretical Physics Vol. 3,
    Pergamon, London/Paris (1958).
\bibitem{LK} I. M. Lifshitz and Yu. Kagan, Zh. Eksp. Teor. Fiz.
    {\bf 62}, 385 (1972) [Sov. Phys. -- JETP {\bf 35}, 206 (1972)].

\end{references}
\end{document}